\documentclass[12pt]{article}
\usepackage[pdftex]{graphicx}
\usepackage{hyperref}
\usepackage{amsmath}
\usepackage{amssymb}
\renewcommand{\title}[1]{%
    \bigskip%
    \begin{center}%
    \Large\bf #1%
    \end{center}%
    \vskip .2in}

\renewcommand{\author}[1]{%
    {\begin{center}
    #1
    \end{center}}}

\begin{document}

\title{\bf{Hamiltonian Formulation of Higher Rank Symmetric  Gauge Theories}}

\centerline{{Rabin Banerjee}\footnote {rabin@bose.res.in}}
\bigskip

\centerline{S. N. Bose National Centre 
for Basic Sciences
, JD Block, Sector III, Salt Lake City,} 
\centerline{Kolkata -700 106, India }

\vskip 1cm
\begin{abstract}
\noindent 
Recent discussions  of fractons have evolved around higher rank symmetric  gauge theories with emphasis on the role of Gauss constraints. This has prompted the present study where a detailed hamiltonian analysis of such theories is presented. Besides a general treatment, the traceless scalar charge theory is considered in details. A new form for the action is given which, in $2+1$ dimensions, yields area preserving diffeomorphisms. Investigation of global symmetries reveals that this diffeomorphism invariance induces a noncommuting charge algebra that gets exactly mapped to the algebra of coordinates in the lowest Landau level problem. Connections of this charge algebra to noncommutative fluid dynamics and magnetohydrodynamics are shown.
\end{abstract}

\section{Introduction }
Higher rank antisymmetric tensor gauge gauge theories, also called $p$-form theories, is an old subject with many ramifications. Recently, another class of higher rank tensor gauge theories have come in the spotlight but, contrary to the $p$- form theories, the gauge fields here are symmetric in the tensor indices \cite{P1, P2, DRH, S}. Their study has led to the discovery and understanding of a new class of topological matter called `fractons', whose properties and applications have been analysed extensively \cite{C, NH,  PCY, QRH, SS, G1, Umang}.  A striking feature is the constrained mobility of these fractons. The excitations are either immobile or, they move in subdimensional spaces; i.e. spaces of dimensionality lower than in which they were formulated. On the other hand, composites of elementary fractons may move freely. These features are a consequence of the existence of nonstandard conservation laws. Contrary to the usual Gauss constraint in Maxwell theory that involves a single derivative of the electric field, here it involves two or more derivatives depending on the rank of the tensor gauge theory. The higher derivative Gauss constraint leads to the conservation of the electric charge (which is the usual conservation law) along with that of the dipole moment and other higher moments of charge distribution. Conservation of the dipole moment now renders a single charge stationary although it allows for partial motion of dipoles. Similarly, other conservation laws have other implications.

It is clear that the Gauss constraint plays a pivotal role in the physical understanding of fractonic excitations. Also, it is crucial in the construction of such higher rank theories. The presence of higher derivatives gives considerable leeway in the construction of electric and magnetic fields, the associated Gauss constraint and the higher rank gauge theory itself \cite{P1, P2}. The general approach is to postulate the basic symplectic structure and the transformation laws from which the Gauss constraint is guessed. From these results, the lagrangian is obtained by inspection \cite{SPP}.  However, it is possible to have a different symplectic structure with a different Gauss constraint to generate identical transformations. In other words, it is desirable, if not essential, to have a structured algorithm from which the various expressions follow. 

Motivated by these possibilities we look for a systematic hamiltonian formulation of higher rank symmetric tensor gauge theories. An overview of the traceless scalar charge theory, which is a prototype of these higher rank theories, is provided in section 2 that illuminates some of the problems and caveats. We next discuss, in section 3, the scalar charge theory, pointing out the differences and similarities with the earlier example. Here (section 3.1) the matter sector is considered in some details, analysing the various global symmetries that yield the fractonic conservation laws. Especially, the results of Seiberg \cite{S} on global symmetries are extended to include another  conservation law besides charge and dipole moment. In section 4 a new lagrangian is suggested that governs the dynamics of a traceless scalar charge theory. This is a mixed system containing both first and second class constraints. While the second class set is eliminated by computing the relevant Dirac brackets, the first class constraints are used to define the gauge generator.  Eventually the first class constraints are also eliminated by a suitable gauge choice that may be interpreted as the analogue of the radiation gauge in usual gauge theories. The resulting gauge fixed brackets are compared with the radiation gauge brackets in Maxwell theory. So far the analysis was in any dimensions. We specialise to $(2+1)$ dimensions in section 5. A specific change of variables is done that solves the traceless constraint $A_{ii}=0$ by expressing the original field $A_{ij}$ in terms of a traceful field, introducing a length scale. A complete hamiltonian analysis reveals the structure of constraints and the gauge generator. It is found  to generate area preserving diffeomorphisms in linearised gravity. As a consequence of this symmetry, elaborated in section 5.1, the charge algebra is found to be noncommuting. In section 6 we show that the area preserving diffeomorphism is exactly mapped to the lowest Landau level problem. The length scale introduced in the change of variables mentioned earlier is identified with the constant magnetic field of the Landau problem. Further, we also show in section 6.1 that the charge algebra is identical to that found in noncommutative fluid dynamics or in magnetohydrodynamics in the presence of a strong magnetic field, examples that mimic the physics of the lowest Landau level problem. Conclusions are given in section 7.

\smallskip

\section{Overview of the traceless scalar charge theory}

 We give here an overview of the traceless scalar charge theory \cite{P1, P2}, basically for a couple of reasons. First, it is a prototype of higher rank tensor gauge theories recently considered in the literature which serves to highlight some of the caveats in the theoretical analysis.  Secondly, it is this particular example that will be treated exhaustively in our paper. These higher rank tensor gauge theories are usually constructed in analogy with the Maxwell theory. The traceless scalar charge theory, for instance,  is defined by a gauge potential $A_{ij}$, which is a symmetric rank 2 tensor $A_{ij}=A_{ji}$, and its conjugate momenta which is the electric field, $E_{ij}$, that is also symmetric.\footnote{Latin indices denote only spatial coordinates, $i,j,..= 1,2,3...d$, while Greek indices will denote  spacetime coordinates. A bar over a variable $(\bar x)$ denotes spatial coordinates only.} These variables satisfy the usual Poisson algebra,
 \begin{equation}
  \{A_{ij}(\bar{x}), A_{kl}(\bar{y})\}= \{E_{ij}(\bar{x}), E_{kl}(\bar{y})\}=0 \,\,;\,\,\{A_{ij}(\bar{x}), E_{kl}(\bar{y})\} = \frac{1}{2}\Big(\delta_{ik}\delta_{jl}+ \delta_{il}\delta_{jk}\Big) \delta(\bar {x}-\bar{y})
  \label{poisson}
 \end{equation}
 The Gauss constraint is an analogue of the usual Maxwell one with an extra derivative but there is another constraint, the tracelessness of the electric field, which has no analogue in the Maxwell theory. These are given by,
 \begin{eqnarray}
 G_1=\partial_i\partial_jE_{ij} &\approx&\rho \cr
 G_2=E_{ii}&\approx& 0
 \label{constraints}
 \end{eqnarray}
 where $\rho$ is the charge. The constraints are implemented weakly in the sense of Dirac \cite{Dirac}. This means they cannot be put directly inside the brackets (\ref{poisson}) (in that case there is a contradiction as may be easily checked from the last relation in (\ref{poisson})), but only after the complete algebra has been computed.
 These constraints lead to the {\it{weak}} conservation of the charges,
 \begin{equation}
     \int_{space}\rho( x)\,\,;\, \int_{space} \bar x\rho( x)\,\,;\, \int_{space}  \bar x^2\rho( x)\label{charges}
 \end{equation}
 While the first one yields the usual charge conservation, the second implies that a charge is immobile while the last indicates that a dipole can move only normally to the dipole moment. Incidentally the last one is a consequence of the tracelessness of the electric field. But, as we shall show, this is not essential and may be obtained even if this condition does not hold.
 
 It is seen that the constraints (\ref{constraints}) generate the following gauge transformations on the potentials,
 \begin{equation}
     A_{ij}\rightarrow A_{ij} + \partial_i\partial_j \alpha + \delta_{ij}\beta
     \label{gaugetransformation}
 \end{equation}
 where $\alpha, \beta$ are the gauge parameters. 
 One possible way to construct a lagrangian invariant under the above transformations is to write a first order form by inspection that would generate the above constraints and from which the hamiltonian is easily read-off. This is the approach adopted in \cite{SPP} where the lagrangian is given by,
 \begin{equation}
   L(E, \phi, \theta, A)= E^{ij}\Big(-\partial_tA_{ij} +\partial_i\partial_j\phi + \delta_{ij}\theta\Big) -\frac{1}{2}\Big(E_{ij}^2 + B_{ij}^2\Big)  
   \label{firstorder}
 \end{equation}
 where $B_{ij}$ is the magnetic field defined in three spatial dimensions as,
 \begin{equation}
     B_{ij}= \frac{1}{2}(\epsilon_{iab} \partial_a A_{bj} + (i\leftrightarrow j)
     \label{magnetic}
 \end{equation}
which is also symmetric and traceless, thereby rendering it gauge invariant,
\begin{equation}
    \{G_i , B_{kl}\} = 0 \,\, \, (i=1, 2)
    \label{magneticinv}
\end{equation}
 The constraints (\ref{constraints}) (for the source free case) are now implemented by the multipliers $\phi$ and $\theta$ while the positive definite terms involving the electric and magnetic fields are identified with the hamiltonian. The equation for $E_{ij}$ defines the electric field,
 \begin{equation}
     E_{ij}= -\partial_tA_{ij} +\partial_i\partial_j\phi + \delta_{ij}\theta
     \label{electric}
 \end{equation}
 Contrary to the magnetic field the electric field in not manifestly tracelees although it is manifestly symmetric. This lack of tracelessness signals a possible caveat in the formulation. Indeed, the imposition of the constraints (\ref{constraints}) on (\ref{electric}) yields certain relations among the variables $(A_{ij})$ and the multipliers $(\phi, \theta)$, the implications of which are not clear and have not been discussed. 
 
 An alternative way to handle the problem was suggested in \cite{Umang} where the second constraint in (\ref{constraints}) was taken along with a new constraint, $A_{ii}\approx 0$. It is now possible to strongly implement the constraints so that $E_{ii}= A_{ii}= 0$ by using Dirac brackets instead of Poisson brackets. This is, however, a purely algebraic manipulation that does not illuminate the dynamical origin of the constraint (\ref{constraints}). The $A-E$ bracket now gets modified. One can compute this by using the usual Dirac procedure. However, in this case it can be readily derived by noting that the constraints are algebraic (there are no differential  operators) so that any correction to the Poisson bracket must be algebraic. Recalling the symmetric nature of both $A_{ij}$ and $E_{ij}$ we find,
 \begin{equation}
     \{A_{ij}(\bar{x}), E_{kl}(\bar{y})\}^* = \frac{1}{2}\Big(\delta_{ik}\delta_{jl}+ \delta_{il}\delta_{jk} - \frac{2}{d} \delta_{ij} \delta_{kl}\Big) \delta(\bar {x}-\bar{y})
     \label{dirac}
 \end{equation}
 The correction term to the Poisson bracket (\ref{poisson}) emerges since the result must be zero for the choices $i=j$ and $k=l$ so that $A_{ii}=0$ and $E_{kk}=0$ are valid. The above modified bracket (denoted by a star) is the relevant Dirac bracket.
 
 After this partial gauge fixing there survives only the higher derivative Gauss constraint, given by the first relation in (\ref{constraints}). In view of the algebra (\ref{dirac}), the Gauss constraint (\ref{constraints}) generates the gauge transformation,
 \begin{equation}
    A_{ij}\rightarrow A_{ij} + \partial_i\partial_j\alpha -\frac{1}{d}\delta_{ij} \partial^2\alpha
    \label{newtransformation}
 \end{equation}
 
 It is useful to make a simple consistency check. Since $A_{ii}=0$ (that was the gauge fixing condition), its variation must vanish; i.e. $\delta A_{ii}=0$. This holds, as may be easily seen from (\ref{newtransformation}).
 
 A gauge invariant lagrangian was also suggested, based on inspection. The electric and magnetic fields were introduced as,
 \begin{eqnarray}
 E_{ij} &=& \partial_i\partial_j A_0 - \frac{1}{d}\delta_{ij}\partial^2A_0 - \partial_t A_{ij}\cr
 B_{ij} &=& -\partial_i\partial_k A_{jk} + \partial_j\partial_k A_{ik}
 \label{fields}
 \end{eqnarray}
 and the lagrangian was written as,
 \begin{equation}
 L= c_1 E_{ij}^2 - c_2 B_{ij}^2
 \label{lag}
 \end{equation}
 However it may be verified easily that this lagrangian {\footnote{Note that for relativistic theories the coefficients $c_1, c_2$ must be identical. This is not required for nonrelativistic theories, although we take them to be equal for simplifying the algebra. The analysis is easily extendable for different $c_1$ and $c_2$, without any basic alteration of results.}} neither yields the Gauss constraints (\ref{constraints}) nor does it yield the symplectic structure leading to the Dirac brackets (\ref{dirac}). This is explicitly shown at the end of section 3, below (25) onward. 
 
 It is thus seen that simply by postulating a set of Gauss constraints might not lead to the construction of a consistent and viable action formulation. Indeed by just giving the Gauss constraints and the gauge transformations, we are already admitting to a specific symplectic structure which the action should reveal. But this need not be achieved as we just saw.  The role of the constraint $E_{ii}\approx 0$, which is algebraic in nature, is also unclear. It does not appear from an analysis of  the lagrangian (\ref{lag}) and, where it appears (see (\ref{firstorder}), it leads to ambiguities. It is thus necessary to develop a systematic formulation where these and other issues are clarified. This is the object of the next section.
 
 We conclude this section by commenting on the structure of the lagrangian (\ref{lag}). Because of the higher derivative nature the form of the electric and magnetic fields is not unique. It is of course possible to carry out the hamiltonian analysis once their explicit forms are known. The constraint structure would change leading to different transformation laws brought about by a change in the Gauss law.
 
 \section{The scalar charge theory}
 
 The  higher derivative theories introduced in the previous section were motivated by the Maxwell theory.
 The principal difference from the standard Maxwell theory is the presence of higher derivatives which gives considerable freedom in defining gauge invariant electric and magnetic fields and hence in the construction of the lagrangian itself.
 
 In this section we present a detailed analysis of  such a theory- the scalar charge theory. We first discuss the free theory and later consider the implications of coupling with sources. The lagrangian is defined, exactly in analogy with the Maxwell theory, by,
 \begin{equation}
     L= \frac{1}{2}\Big( E_{ij}^2 - B_{ij}^2\Big)
     \label{lagrangian}
 \end{equation}
 where the electric and magnetic fields have been introduced in (\ref{fields}).
 The canonical momenta are given by,
 \begin{equation}
     \pi_0=\frac{\partial L}{\partial \dot A_0} =0\,\,\,;\,\,\,  \pi_{ij}=\frac{\partial L}{\partial \dot A_{ij}} = -E_{ij}
     \label{momenta}
 \end{equation}
with the over-dot indicating differentiation with respect to time.

 Since it involves time derivatives, only $\pi_{ij}$ is a genuine momenta. The other one is a constraint, a primary constraint,
 \begin{equation}
   \Phi_1=\pi_0\approx 0
   \label{primary}
 \end{equation}
 In order to find the secondary constraints, if any, time conservation of the primary constraint is required. To do that, we have to write the total hamiltonian \cite{Dirac}, defined from the canonical hamiltonian as,
 \begin{eqnarray}
 H_T &=& H_c + \int_{space} \lambda \Phi_1\cr
      &=& \int_{space} \Big(\pi^0 \dot A_0 + \pi_{ij} \dot A_{ij} - L +\lambda \Phi_1\Big)\cr
    &=& \int_{space}  \Big(\frac{1}{2}(\pi_{ij}^2 + B_{ij}^2) +\pi_{ij}(\partial_i\partial_j - \frac{1}{d}\delta_{ij} \partial^2)A_0 +\lambda \Phi_1\Big) 
    \label{totalham}
 \end{eqnarray}
 where $\lambda$ is a multiplier enforcing the primary constraint. The Poisson algebra among the basic variables is given by (\ref{poisson}) and,
 \begin{equation}
     \{A_0(x), \pi^0(y)\}= \delta(\bar x - \bar y)
     \label{mpoisson}
 \end{equation}
 Time conserving the primary constraint $\{\Phi_1, H_T\}\approx 0$ immediately yields the secondary constraint, which is the Gauss constraint,
 \begin{equation}
    \Phi_2= (\partial_i\partial_j - \frac{1}{d} \delta_{ij}\partial^2)\pi_{ij}\approx 0
    \label{gauss}
    \end{equation}
    
    No further constraints are generated by this iterative process
since $\{\Phi_2, H_T\}= 0$. As the constraints are involutive,
\begin{equation}
    \{\Phi_i, \Phi_j\}=0 \,\,;\,\, i,j=1,2
\end{equation}
 these are first class. The system is thus a clean example of a gauge theory. The gauge generator, following Dirac's conjecture, is a linear combination of all first class constraints of the theory. Thus, it is given by,
 \begin{equation}
    \Omega= \int_{space} (\alpha_1(x)\Phi_1(x) + \alpha_2(x)\Phi_2(x))
    \label{generator}
 \end{equation}
 where $\alpha_1, \alpha_2$ are the gauge parameters. However these are not independent. The number of independent parameters is given by the number of independent primary first class constraints, which is one in this case. There is a set of equations from which the relation between the parameters can be obtained \cite{BRR1, BRR2}. In this case, however, we find this by an alternative method. 
 
 The gauge generator generates the following transformations on the fields,
 \begin{eqnarray}
 \delta A_0&=&\{A_0, \Omega\}= -\alpha_1\cr
 \delta A_{ij}&=& \{A_{ij}, \Omega\}= (\partial_i\partial_j - \frac{1}{d} \delta_{ij}\partial^2)\alpha_2\cr
 \delta \pi_{ij}&=&-\delta E_{ij} = 0
 \label{gt0}
 \end{eqnarray}
 with the last relation merely showing the gauge invariance of the electric field. We now take the variation of the fields appearing on either side of the first equation in (\ref{fields}), using (\ref{gt0}). We find,
 \begin{equation}
     0= -\partial_i\partial_j\alpha_1+\frac{1}{d}\delta_{ij}\partial^2 \alpha_1- (\partial_i\partial_j - \frac{1}{d} \delta_{ij}\partial^2)\dot\alpha_2
 \end{equation}
 from which we immediately obtain,
 \begin{equation}
     \alpha_1=-\dot\alpha_2
 \end{equation}
 Renaming $\alpha_2$ as  $\alpha$, we obtain the following transformations under which (\ref{lagrangian}) is invariant,
 \begin{equation}
     \delta A_0=\dot\alpha\,\,;\,\, \delta A_{ij}= (\partial_i\partial_j - \frac{1}{d} \delta_{ij}\partial^2)\alpha\label{transfm}
 \end{equation}
 Although the transformation for $A_{ij}$  reproduces the result (\ref{newtransformation}), there are crucial differences. The structure of the Gauss constraint in the two cases (\ref{constraints}), (\ref{gauss}) is distinct and so is the  algebra among the basic variables. The two differences cancel to yield the same result. This reinforces the necessity to carry out a systematic analysis by starting from a specific lagrangian instead of simply postulating certain transformations.
 
 In this presentation the gauge choice $A_{ii}\approx 0$ cannot even be done, let alone reproducing the result (\ref{dirac}). This is because $A_{ii}$ is gauge invariant, having a vanishing algebra with the Gauss constraint (\ref{gauss}),
 \begin{equation}
     \{ A_{ii}\, , \, \Phi_2\} = 0
 \end{equation}
 This is the physical reason. Algebraically, the matrix formed by the Poisson brackets involving the complete set of constraints- the Gauss constraint and the gauge condition -  becomes noninvertible so that the Dirac brackets cannot be defined. The same conclusion holds if we started from (\ref{lag}) instead of (\ref{lagrangian}) (see footnote 3).
 
 It is actually possible to prove that, in this theory,  there is no gauge choice that yields the symplectic structure (\ref{dirac}). This quite general  statement further bolsters the observation  made below (\ref{lag}). Any valid gauge choice would lead to Dirac brackets that satisfy the strong imposition  of both the Gauss constraint (\ref{gauss}) as well as the gauge condition. Assuming that there is a gauge choice that yields (\ref{dirac}), then it must satisfy the condition,
 \begin{equation}
     \{A_{ij}(\bar{x}), (\partial_k\partial_l - \frac{1}{d} \delta_{kl}\partial^2)E_{kl}(\bar{y})\}^* = 0 
     \label{ref}
 \end{equation}
 noting that the canonical momenta and the electric field just differ by a sign.
 This does not hold as may be easily seen by applying the differential operator on the right side of (\ref{dirac}). We find a contradiction,
 \begin{equation}
      \{A_{ij}(\bar{x}), (\partial_k\partial_l - \frac{1}{d} \delta_{kl}\partial^2)E_{kl}(\bar{y})\}^*= 2\Big(\partial_i\partial_j - \frac{1}{d}\partial^2\delta_{ij}\Big)\delta(x-y)\neq 0
      \label{ref1}
 \end{equation}
 Thus the algebra (\ref{dirac}) is untenable.
 
 To complete the picture, we choose an appropriate gauge and compute the symplectic structure. A valid gauge choice is given by,
 \begin{equation}
     \partial_i\partial_j A_{ij}\approx 0\label{ref2}
 \end{equation}
 With such a choice the Dirac brackets are given by,{\footnote{Since a detailed computation of such brackets is provided in section 4, here the result is just given.}}
 \begin{equation}
     \{A_{ij}(\bar{x}), E_{kl}(\bar{y})\}^* =\Big[ \frac{1}{2}(\delta_{ik}\delta_{jl}+ \delta_{il}\delta_{jk} )- \frac{d}{1-d}\Big(\frac{1}{d}\delta_{ij}- \frac{\partial_i\partial_j}{\partial^2}\Big)\frac{\partial_k\partial_l}{\partial^2}\Big] \delta(\bar {x}-\bar{y})
 \end{equation}
 It is verified that this structure is compatible with the strong imposition of both the Gauss constraint (\ref{gauss}) as well as the gauge condition (\ref{ref2}).
 
 \subsection{The matter sector}

 Let us now introduce sources with $J^0$ and $J_{ij}$ coupling with $A_0$ and $A_{ij}$ respectively. We first discuss the pure matter sector, specifically the global symmetries \cite{S},  and then follow it up by the complete theory. Gauge invariance under (\ref{transfm})  implies the conservation law,{\footnote{Notation: temporal indices are denoted by 0 while Latin indices indicate space, the two are combined by using Greek indices. Temporal indices change sign on lowering or raising, spatial ones do not. Nonrelativistic physics is being discussed which is made transparent in the last equality of (\ref{m0}).}}
 \begin{equation}
   \partial_0 J_0 -  (\partial_i\partial_j - \frac{1}{d} \delta_{ij}\partial^2)J_{ij} = \partial_t\rho -  (\partial_i\partial_j - \frac{1}{d} \delta_{ij}\partial^2)J_{ij} = 0
   \label{m0}
 \end{equation}
 This result ensures the conservation of the three charges (\ref{charges}) without any restriction on the sources.  This is shown in some details using the global symmetries. It has the usual global symmetry,
 \begin{equation}
     \partial_\mu j^\mu=0\,\,\leftrightarrow \partial_0 j_0-\partial_ij_i = 0
     \label{m1}
 \end{equation}
 where,
 \begin{equation}
     j_0=J_0 \,\,; \,\, j^i= \partial_j J^{ji}-\frac{1}{d}\partial_i J \,\,\, (J=J_{ii})
     \label{m2}
 \end{equation}
 and $J$ is the trace of $J_{ij}$,{\footnote{The trace of other variables are denoted similarly, $B_{ii}=B$ etc.}} as shown above. This leads to the usual conserved charge,
 \begin{equation}
     Q=\int_{space} j_0=\int_{space} J_0
     \label{m3}
 \end{equation}
 
 Further, it has a vector global symmetry with currents,
 \begin{equation}
    j_0^i=x^i J_0\,\,;\,\, j^{ij}= x^j\partial_kJ^{ki} - \frac{1}{d}x^j\partial^i J +\frac{1}{d}\delta^{ij} J -J^{ij}
    \label{m4}
 \end{equation}
 which yields the conservation law,
 \begin{equation}
     \partial_0 j_0^i - \partial_j j^{ji}=0
     \label{vectorglobal}
 \end{equation}
 that may be verified from (\ref{m0}). The conserved charge here is,
 \begin{equation}
     Q^i=\int_{space} j_0^i= \int_{space} x^i J_0
     \label{vectorcharge}
 \end{equation}
 
 Finally, there is another scalar charge which is different from the usual one (\ref{m3}). The global symmetry is here defined by the currents,
 \begin{equation}
     j_0=x^2 J_0\,\,\,;\,\,\, j^i=x^2\Big(\partial_j J^{ji} - \frac{1}{d}\partial^i J\Big) - 2x_j J^{ji} +\frac{2}{d} x^i J
     \label{quadratic}
 \end{equation}
 and satisfies an identical conservation law as (\ref{m1}), which is verified by using (\ref{m0}). The corresponding conserved charge is given by,
 \begin{equation}
     Q_0=\int_{space}  j_0 = \int_{space} x^2 J_0
     \label{quadratic1}
 \end{equation}
 The three charges (\ref{m3}, \ref{vectorcharge}, \ref{quadratic1}) are those mentioned in (\ref{charges}).{\footnote{Note that, in usual literature, the conservation of the charge (\ref{quadratic1}) is achieved only if the traceless condition is imposed $J=J_{ii} =0$ \cite{Umang}. This is not necessary here. (See also the discussion below (\ref{charges})).}}
 
 It is possible to extend this analysis for multipole moments. Additional conservation laws would emerge corresponding to these higher moments. It also shows a connection with the  conventional approach using higher rank symmetric tensor fields. More conservation laws follow from the introduction of higher rank gauge fields that couple with corresponding higher rank tensor sources, leading to a generalisation of (\ref{m0}).
 
 Let us next gauge the usual global symmetry (\ref{m0}) and write the complete lagrangian as,
 \begin{equation}
     L=L_0 +  \frac{1}{2}\Big( E_{ij}^2 - B_{ij}^2\Big)+A_0 J_0 +  A_{ij} J_{ij}
     \label{totallag}
 \end{equation}
 where $L_0$ is the contribution from the matter sector and the electric and magnetic fields have been defined in (\ref{fields}).{\footnote{In those cases where the symmetric field $A_{ij}$ can be written in terms of the conventional $U(1)$ field as $A_{ij}= \frac{1}{2}(\partial_i A_j + \partial_j A_i)$, the coupling in (\ref{totallag}) may be expressed as $-A_i\partial_j J_{ij}$, which is equivalent to the discussion in Seiberg \cite{S}. This decomposition is possible if $\delta A_{ij}= \partial_i\partial_j\alpha$. This is not true here as may be seen from (\ref{transfm}).}}
 
 The equations of motion of the gauge fields are given by,
 \begin{eqnarray}
 \partial_i\partial_j E^{ij} -\frac{1}{d}\partial^2 E +J_0 &=& 0\cr
 \partial_0E_{ij} +\partial_l\partial_j B_{li} +\partial_l\partial_i B_{lj} + J_{ij} &=& 0
 \label{gaugeequations}
 \end{eqnarray}
 
 After this gauging, the current of the global symmetry (\ref{m2}), (\ref{m3}) may be corrected by improvement terms such that it trivialises, exactly as happens for the standard $U(1)$ gauge field,
 \begin{eqnarray}
 j_0\rightarrow j_o + \partial_i\partial_j E^{ij} -\frac{1}{d}\partial^2 E &=& 0\cr
 j_i\rightarrow j_i + \partial_0\partial_j E_{ij} + \partial_j\partial^2B_{ji} -\frac{1}{d}\partial_0\partial_i E&=&0
 \label{improved}
 \end{eqnarray}
 where use was made of the equations of motion (\ref{gaugeequations}). As a consistency check, it can be shown that the above currents satisfy the conservation law (\ref{m1}).{\footnote{ Incidentally the currents (\ref{m1}) and the charge (\ref{m2}) are defined only modulo the improvement terms \cite{S},
 \begin{eqnarray}
 j_0&\rightarrow& j_0 + \partial_i X^i\cr
 j_i&\rightarrow& j_i + \partial_0X_i + \partial_j Y_{ji}\,\,;\,\,Y_{ji}=-Y_{ij}
   \label{basicimp}  
 \end{eqnarray}
 In the present example, $X_i= \partial_jE_{ij} -\frac{1}{d}\partial_iE,\,\, Y_{ji}= \partial^2B_{ji}$, which yields (\ref{improved}).}}
 
 \section{The traceless scalar charge theory}
  
 In this section we analyse the traceless scalar charge theory. Apart from comparing with previous approaches and results, we use these findings to subsequently discuss diffeomorphism symmetry from which the physics of the lowest Landau problem emerges naturally.
 
 The lagrangian is defined by,
 \begin{equation}
      L= \frac{1}{2}\Big( E_{ij}^2 - B_{ij}^2\Big) + \alpha A_{ii}
     \label{lagrangian1}
 \end{equation}
 which, as far as we are aware, was not considered earlier. The difference from the lagrangian (\ref{lagrangian}) of the scalar charge theory is the presence of the last term that enforces the tracelessness of the tensor gauge field.
 
 The canonical momenta are given by,
 \begin{equation}
     \pi_{ij}=-E_{ij};\,\,\,\pi_0=0;\,\,\,\pi_\alpha=0;\,\,\,\pi_{ii}=0
     \label{momenta1}
 \end{equation}
 Only the first one is a true momentum while the others are all (primary) constraints which have to be implemented weakly,
 \begin{equation}
     \Phi_1=\pi_0\approx 0,\,\,\, \Phi_2=\pi_\alpha\approx 0,\,\,\, \Phi_3=\pi_{ii}=-E_{ii} \approx 0\label{tracelessconstraints}
 \end{equation}
 To get the secondary constraints we have to first write the total hamiltonian,
 \begin{equation}
     H_T= \int_{space}  \Big(\frac{1}{2}(\pi_{ij}^2 + B_{ij}^2) +\pi_{ij}(\partial_i\partial_j - \frac{1}{d} \delta_{ij}\partial^2)A_0-\alpha A_{ii} +\chi_i\Phi_i\Big) ;\,\, i=1, 2, 3
     \label{trconstraints}
 \end{equation}
 where $\chi_i$ are the multipliers enforcing the constraints $\Phi_i$. Time conservation of the primary constraints $\Phi_1$ and $\Phi_2$ yield further constraints,
 \begin{equation}
     \Phi_4=\{\Phi_1, H_T\}=(\partial_i\partial_j - \frac{1}{d} \delta_{ij}\partial^2)\pi_{ij}\approx 0,\,\,\Phi_5 =\{\Phi_2, H_T\}=A_{ii}\approx 0
     \label{newconsts}
 \end{equation}
 The $\Phi_3$ constraint does not generate any new constraint since, along with $\Phi_5$, it forms a second class pair.  The other three $\Phi_1, \Phi_2,  \Phi_4$ are first class since their algebra closes with {\it{all}} the constraints. We mention in passing that, contrary to usual approaches, here $E_{ii}\approx 0$ is not any gauge generator and neither is $A_{ii}\approx 0$  any gauge fixing condition. This pair of second class constraints is eliminated by calculating the relevant Dirac brackets and the answer was given in (\ref{dirac}). 
 
 It is thus clear that the formulation of a dynamical model for the traceless theory is nontrivial. If we first perform a canonical (hamiltonian) analysis and then constrain by imposing the traceless condition as a gauge fixing condition, we fail, as shown in the earlier section. If, on the other hand, we first impose the traceless condition by hand in the lagrangian and then perform the canonical analysis, we succeed. This is a typical example where canonical analysis and imposition of constraints do not commute and is a well known feature in constrained dynamics.   Not only that, in the latter case we reproduce the algebra (\ref{dirac}).  Significantly, in the former approach, the constraint $E_{ii}=0$, which is an essential companion of $A_{ii}=0$, never appears.

 After the strong imposition of the second class constraints, the only physically relevant first class constraints are given by,
 \begin{equation}
     \pi_0\approx 0\,,\,\, \Pi=\partial_i\partial_j \pi_{ij}\approx 0
     \label{fcc}
 \end{equation}
 The gauge generator is now given by,
 \begin{equation}
 G=\int_{space}  \Big( \lambda_0\pi_0 + \lambda \,\partial_i\partial_j\pi_{ij}\Big)
 \label{generator1}
 \end{equation}
where $\lambda_0$, $\lambda$ are gauge parameters. Since there is only one primary first class constraint $(\pi_0\approx 0)$, there is only one independent gauge parameter. Using the method discussed before we find that $\lambda_0=-\dot\lambda$. The fields $A_0$ and $A_{ij}$ transform exactly like (\ref{transfm}).

At this point the consistency check discussed around (\ref{newtransformation}) is recalled. This also holds here. If, on the other hand, the scalar charge theory was taken with the transformation (\ref{gaugetransformation}), it would be incompatible with $\delta A_{ii}=0$. This shows the need for making cross checks in the consistency of the formulation.
 
 The structure of the constraints shows a close resemblance to the Maxwell theory. This may be pushed further if we perform the  gauge fixing,
 \begin{equation}
   \Lambda= \partial_i\partial_j A_{ij}\approx 0
    \label{radgauge}
 \end{equation}
 which may be considered the analogue of the radiation gauge $\partial_i A_i\approx 0$ for the Maxwell theory. Together, the Gauss constraint and the gauge condition form a second class pair m of constraints and are eliminated by computing relevant Dirac brackets. The $A-\pi \,(A-E)$ algebra is modified. The relevant Dirac bracket is defined as,
 \begin{equation}
     \{A_{ij}(\bar x), \pi_{kl}(\bar y)\}^{**}= \{A_{ij}(\bar x), \pi_{kl}(\bar y)\}^* - \int d\bar z d\bar z_1\{A_{ij}(\bar x), \Pi(\bar z)\}^* \{\Pi(\bar z), \Lambda(\bar z_1)\}^{*-1}\{\Lambda(\bar z_1), \pi_{kl}(\bar y)\}^*
 \end{equation}
 where the $**$ indicates the final Dirac bracket which is computed in terms of the $*$ bracket, which is the Dirac bracket derived at the first stage of the analysis when the original second class constraints were eliminated. Effectively the $*$ bracket takes 
 over the role of the Poisson bracket in the usual definition of the Dirac bracket. The inverse that appears above is the inverse of the star bracket involving the constraints. Incidentally the relevant $*$ bracket (i.e. the first level Dirac bracket) has been defined in (\ref{dirac}). After some algebra the final result is obtained,
 \begin{equation}
     \{A_{ij}(\bar x), \pi_{kl}(\bar y)\}^{**}= \{A_{ij}(\bar x), \pi_{kl}(\bar y)\}^* -\frac{d}{d-1}\Delta_{ij}\Delta_{kl} \delta(\bar x-\bar y)\label{finaldirac}
 \end{equation}
 where,
 \begin{equation}
     \Delta_{ij} = \Big(\frac{\partial_i\partial_j}{\partial^2} - \frac{\delta_{ij}}{d}\Big)
 \end{equation}
 and ensures the vanishing of the Dirac brackets,
 \begin{equation}
     \{\partial_i\partial_jA_{ij}, \pi_{kl}\}^{**}= \{A_{ij}, \partial_k\partial_l \pi_{kl}\}^{**} =0
 \end{equation}
 This shows that the constraints  are now implemented strongly by the final Dirac brackets, so that, $\Pi= 0, \Lambda= 0$. Of course these brackets also satisfy $A_{ii}= \pi_{ii}=0$, which were the second class pair of constraints before any gauge fixing was done.
 
 It is useful to recall the example of the Maxwell theory where the Gauss constraint $\partial_i\pi_i\approx 0$ is fixed by the radiation gauge constraint $\partial_iA_i\approx 0$ and the expression for the Dirac bracket is,
 \begin{equation}
     \{A_i(\bar x), \pi_j(\bar y)\}=\Delta_{ij}^T(\bar x-\bar y)
 \end{equation}
 where the transverse delta function is defined as,
 \begin{equation}
     \Delta_{ij}^T(\bar x-\bar y)=\Big(\delta_{ij}-\frac{\partial_i\partial_j}{\partial^2}\Big)\delta(\bar x-\bar y)
 \end{equation}
 satisfying,
 \begin{equation}
     \partial_i\Delta_{ij}^T=\partial_j\Delta_{ij}^T=0
 \end{equation}
 As we see the structure in the present case (\ref{finaldirac}) is much more involved than the Maxwell example. The reasons are twofold: the presence of higher order derivatives and the occurrence of the traceless constraints $A_{ii}\approx 0, \pi_{ii}\approx 0$ which do not have any analogue in the Maxwell theory.
 
 \section{Scalar charge theory in (2+1) dimensions and diffeomorphism symmetry}
 
 Having discussed the issue of gauge fixing, we reconsider the theory (\ref{lagrangian1}) where the second class constraints were eliminated but the important Gauss constraint $\Pi\approx 0$ (\ref{fcc}) remained as a first class constraint. If we now specialize to $(2+1)$ dimensions (i.e. d=2) we find interesting physical consequences. One of these is discussed here where we are able to construct a theory that has diffeomorphism symmetry which may be interpreted as a theory of linearized gravity.
 
 Since the traceless constraint $A_{ii}=0$ in (\ref{lagrangian1}) is strongly imposed, it is possible to solve for it directly in terms of another symmetric, but not traceless, second rank field in the manner,
 \begin{equation}
     A_{ij}= -\frac{1}{4l^2}\Big(\epsilon_{kj}h_{ki}+\epsilon_{ki}h_{kj}\Big)
     \label{newlagrangian}
 \end{equation}
 The above construction ensures the symmetric and traceless nature of $A_{ij}$, using only the symmetric nature of $h_{ij}$.{\footnote{The inverse relation involves the trace of $h_{ij}$,
 \begin{equation}
       h_{ij}- \frac{1}{2}\delta_{ij} h = -l^2\Big(\epsilon_{jk}A_{ki}+\epsilon_{ik}A_{kj}\Big) 
     \label{newlagrangianinverse}
 \end{equation}
 }}
 The length scale $l$ is introduced for dimensional reasons. Later, it will acquire a greater significance. Substituting in (\ref{lagrangian1}) we obtain the new lagrangian expressed in terms of the $h$ field. The result is,
 \begin{equation}
      L= \frac{1}{2}\Big( E_{ij}^2 - B_{ij}^2\Big) 
     \label{lagrangian2}
 \end{equation}
 where the electric and magnetic fields, computed from (\ref{fields}), are given by,
 \begin{eqnarray}
 E_{ij}&=&\partial_i\partial_jA_0-\frac{1}{2}\delta_{ij}\delta^2A_0+\frac{1}{4l^2}\Big(\epsilon_{ki}\partial_th_{kj}+\epsilon_{kj}\partial_th_{ki}\Big)\cr
 \epsilon_{ij}B_{ij}&=&\frac{1}{l^2}\Big(\partial_i\partial_jh_{ij} - \frac{1}{2}\partial^2h_{ii}\Big)
 \label{newfields}
 \end{eqnarray}
 Since the magnetic field has only one component $B_{12}$, it is convenient to express it in the way done above.
 
 We now perform a canonical analysis of the above model. The canonical momenta are defined by,
 \begin{equation}
     \pi^0=\frac{\partial L}{\partial \dot A_0} = 0\,;\,\,\pi_{ij}= \frac{\partial L}{\partial \dot h_{ij}}= \frac{1}{4l^2}\Big(\epsilon_{ik} E_{kj} + \epsilon_{jk} E_{ki}\Big)\label{newmomenta}
 \end{equation}
 where the electric field is given in (\ref{newfields}).
 A useful identity that will be used later on follows,
 \begin{equation}
     \pi_{ij}^2 = \frac{1}{4l^4} E_{ij}^2
     \label{identity}
 \end{equation}
 There is one primary constraint,
 \begin{equation}
     \pi_0\approx 0
     \label{newprimary}
 \end{equation}
 while the other is a true momenta. The total hamiltonian is now found to be,
 \begin{equation}
     H_T=\int_{space} \Big(2l^4 \pi_{ij}^2 + \frac{1}{2} B_{ij}^2 +  A_0 \Phi+\lambda \pi_0\Big)
     \label{newtotalham}
 \end{equation}
 where,
 \begin{equation}
     \Phi= -2l^2\epsilon_{pi}\partial_i\partial_q\pi_{pq}
     \label{newgauss}
 \end{equation}
 and the primary constraint is enforced by the lagrange multiplier $\lambda$. Time conserving the primary constraint yields the Gauss constraint,
 \begin{equation}
     \Phi\approx 0\label{GAUSS}
 \end{equation}
 There are no more constraints since time conservation of the Gauss constraint yields a vanishing result,
 \begin{equation}
     \{\Phi, H_T\}=0
 \end{equation}
 The physical space is defined to be that space which is annihilated by the first class constraints,
 \begin{equation}
     \pi_0\,|\Psi>_P=\Phi\,|\Psi>_P=0
 \end{equation}
 so that the total hamiltonian in the physical subspace, after using the identity (\ref{identity}), simplifies to,
 \begin{equation}
     H_P=\int_{space} \Big(\frac{1}{2}E_{ij}^2 + \frac{1}{2} B_{ij}^2   \Big)
     \label{newtotalham1}
 \end{equation}
 which takes on a familiar look.
 
 \subsection{Volume preserving diffeomorphism symmetry  and linearised gravity}
 
 We next consider the gauge symmetries which will eventually lead to volume preserving diffemorphisms.The generator of the gauge transformations is given, as usual, by a linear combination of the first class constraints,
 \begin{equation}
     G=\int_{space} \Big( \alpha_0\pi_0 + \alpha\Phi\Big)
     \label{generator1}
 \end{equation}
 where $\alpha_0, \alpha$ are the gauge parameters.
 Then the gauge variations are given by,
 \begin{eqnarray}
 \delta A_0&=&\{A_0, G\}= -\alpha_0\cr
 \delta h_{ij}&=& \{h_{ij}, G\}=-l^2\Big( \epsilon_{il}\partial_l\partial_j\alpha +\epsilon_{jl}\partial_l\partial_i\alpha\Big) 
 \label{gt}
 \end{eqnarray}
  where we have used the basic Poisson algebra (\ref{poisson}) to compute the above brackets.   

 Since there is only one primary first class constraint, there is one independent gauge parameter. As done earlier we get the relation between the parameters by 
 taking the variation on either side of the electric field in (\ref{newfields}). As the electric field is linearly related to the canonical momenta (\ref{newmomenta}) it is obviously gauge invariant. Using the transformations (\ref{gt}), we find,
 \begin{equation}
     0= -\partial_i\partial_j\alpha_0 + \frac{1}{2} \delta_{ij}\partial^2\alpha_0 -\partial_i\partial_j\dot\alpha + \frac{1}{2} \delta_{ij}\partial^2\dot\alpha 
 \end{equation}
 which immediately yields $\alpha_0=-\dot\alpha$, so that the gauge generator takes the final form,
 \begin{equation}
     G=\int_{space} \Big( -\dot\alpha\pi_0 + \alpha\Phi\Big)
     \label{generatorfinal}
 \end{equation}
 The transformation on the $A_0$ and $h$ fields may be expressed as,
 \begin{equation}
    \delta A_0= \dot\alpha\,;\,\, \delta h_{ij}= -\partial_i\eta_j - \partial_j \eta_i\,\,;\, \eta_i= l^2\epsilon_{il}\partial_l\alpha
     \label{diffeos}
 \end{equation}
 The second relation is exactly the transformation of a spatial metric under volume preserving diffeomorphisms,{\footnote{The word `volume' is used generically since in two dimensions, which is  relevant here, we have area preserving diffeomorphisms.}} $x_i\rightarrow x_i + \eta_i$ because $\partial_i\eta_i=0$, as seen from (\ref{diffeos}). A similar transformation was discussed recently in \cite{Umang} where the metric was traceless $h_{ii}=0$ leading to unimodular gravity. Our example  is general $(h_{ii}\neq 0)$ and hence  does not have this restriction. 
 
 
 Instead of considering $h_{ij}$ as a gauge field it is possible to interpret it as the linear correction to $g_{ij}$, expanded around a flat background,
 \begin{equation}
  g_{ij}=\delta_{ij} + h_{ij}
  \label{flat}
 \end{equation}
 Under the infinitesimal volume preserving transformation mentioned earlier the metric transforms as,
 \begin{equation}
    \delta_\alpha  g_{ij}= -\Big( \eta^k \partial_k g_{ij} +g_{kj}\partial_i\eta^k  +g_{ik}\partial_j\eta^k\Big) = 
    -l^2 \epsilon^{kl}\Big(\partial_k g_{ij} + g_{kj} \partial_i +g_{ki}\partial_j\Big)\partial_l\alpha
    \label{metrictrans}
    \end{equation}
    where the first piece is the transport term while the other two come from the form variation of the space components of a second rank tensor. If we now substitute (\ref{flat}) in (\ref{metrictrans}) and retain terms in the leading order only, then the result (\ref{diffeos}) is reproduced.
    
    As is known, volume preserving transformations lead to a nonlinear realisation of the symmetry given by,
    \begin{equation}
        [\delta_\eta, \delta_\rho] = \delta_{[\eta, \rho]}
        \label{nonlinear}
    \end{equation}
 where, in the current example,{\footnote{The algebra of transformation in (\ref{nonlinear}) is a genuine commutator. However, the one that occurs in (\ref{nonlinearparameter}) is not because $\eta, \rho$ are classical objects and there is no non-abelian nature linked with them. For a non-abelian gauge theory, for instance, this would be a true commutator involving the algebra of group matrices. The symmetry here is nonlinear and not non-abelian. Thus (\ref{nonlinearparameter}) is interpreted as a Poisson bracket which explains the notation.}}
 \begin{equation}
     [\eta, \rho]_{PB}= l^2\epsilon_{ij} \partial_i\eta \partial_j \rho
     \label{nonlinearparameter}
 \end{equation}
 as may be easily verified by using (\ref{metrictrans}) and the definition of the transformation parameter given in (\ref{diffeos}).
 
 Similar conclusions hold for the transformation of the field $A_0$. By including the transport term, its total variation is obtained from (\ref{diffeos}),
 \begin{equation}
     \delta_\alpha A_0= \partial_t\alpha - \eta^k\partial_kA_0=\partial_t\alpha - l^2\epsilon^{kl}\partial_kA_0\partial_l\alpha
     \label{Afield}
 \end{equation}
 which also satisfies the closure relation (\ref{nonlinear}).
 
 So far we have been discussing generators and transformations related to the gauge sector. For the matter sector, the corresponding operators are the charge density $\rho$ that couples with $A_0$ and the stress tensor $T^{ij}$ that couples with $A_{ij}$. For the charge sector, the change of any matter field $\Psi(x)$ under infinitesimal gauge transformations is given by,
 \begin{equation}
     i\delta_\alpha \Psi(x)= \int d^2y \, [\alpha (y)\rho(y),\, \Psi(x)]
     \label{mattervariation}
 \end{equation}
 If we take the variation of the charge itself by putting $\Psi=\rho$ in the above relation and ensure consistency with the closure relation for the diffeomorphisms (\ref{nonlinear}) then the algebra of charges follows,
 \begin{equation}
     [\rho(\bar x), \,\rho(\bar y)]=il^2\epsilon^{kl}\partial_k\rho(\bar x)\partial_l\delta(\bar x- \bar y)
     \label{chargealgebra}
 \end{equation}
 
 We will now exploit these results to establish a connection with the lowest Landau level problem. To do that we first review the Landau problem.
 
 \section{Connection with Landau problem in presence of a strong magnetic field}
 In order to establish a clean connection with the physics of the lowest Landau level problem, we first consider the lagrangian of a charged particle moving in a plane under the influence of a constant magnetic field $B$ ,
 \begin{equation}
     L= \frac{1}{2}m \dot x_i^2 + e\dot x_i A_i - V\,; \, (i=1, 2)
     \label{landau}
 \end{equation}
 where we have set $c=1$ and work in the radiation gauge $(\partial_i A_i=0)$, so that,
 \begin{equation}
     A_i= \frac{B}{2} \epsilon_{ij}x_j
     \label{gauge}
 \end{equation}
 and $V$ is the potential from which other forces can be derived.
 
 The equation of motion following from the lagrangian is,
 \begin{equation}
     m\ddot x_i = -eB\epsilon_{ij}\dot x_j -\partial_iV
     \label{lorentz}
 \end{equation}
 which is the Lorentz force law.
 
 In the hamiltonian formulation, the conjugate momenta are given by,
 \begin{equation}
     p_i=\frac{\partial L}{\partial \dot x_i}= m\dot x_i + \frac{eB}{2}\epsilon_{ij}x_j
     \label{conmomenta}
 \end{equation}
 The canonical hamiltonian is obtained from the lagrangian by a standard Legendre transformation,
 \begin{equation}
     H= \frac{\pi_i^2}{2m} + V
     \label{lanham}
 \end{equation}
 where $\pi_i$ is the kinematical momenta obtained from the canonical momenta by a minimal substitution,
 \begin{equation}
     \pi_i= p_i - eA_i= p_i - \frac{eB}{2}\epsilon_{ij}x_j
     \label{zero}
 \end{equation}
 
 The projection to the lowest Landau level is achieved in the strong magnetic field case. Then the mass term in (\ref{landau}) can be set to zero, leading to the lagrangian,
 \begin{equation}
     L=\frac{eB}{2} \epsilon_{ij} x_j \dot x_i -V
     \label{LLLL}
 \end{equation}
 This reduction to a first order system enables one to simply read off the brackets without entering into the elaborate Dirac procedure. The canonical pair is $(x_1, eB x_2)$ so that the basic algebra is given by,
 \begin{equation}
     \{x_i, x_j\}=\frac{1}{eB} \epsilon_{ij}
     \label{landaualgebra}
     \end{equation}
     The equation of motion is the same as found from (\ref{lorentz}) by putting $m=0$,
     \begin{equation}
         eB\epsilon_{ij}\dot x_j +\partial_iV=0
         \label{llmotion}
     \end{equation}

 These results are now rederived in the hamiltonian formulation. It is done not merely to establish compatibility but also to provide justification in those examples, one of which will be treated in section 6.1, where a straightforward lagrangian approach is unavailable. The hamiltonian following from (\ref{LLLL}) is,
 \begin{equation}
     H=V
     \label{lham}
 \end{equation}
 It will reproduce the above equation of motion (\ref{llmotion}) provided we take the basic algebra as (\ref{landaualgebra}),
 \begin{equation}
     \dot x_i= \{x_i, V\}= \partial_jV\{x_i, x_j\}=\frac{1}{eB} \epsilon_{ij}\partial_j V
     \label{discussion}
 \end{equation}
 
 Let us now start from the hamiltonian (\ref{lanham}) in the $m\rightarrow 0$ limit. To make the first term meaningful it is necessary to take $\pi_i=0$. One could argue that one might as well take $\pi_i^2=0$. The actual justification for taking $\pi_i=0$ comes from (\ref{conmomenta}) and (\ref{zero}) which shows that for $m=0$ we have $\pi_i=0$.  Now the first term in (\ref{lanham}) has to be interpreted by initially setting the numerator to zero strongly in which case the hamiltonian reduces to the earlier result (\ref{lham}) derived directly from the lagrangian (\ref{LLLL}), so that consistency is retained. Once $\pi_i=0$ strongly, there is a clash among the various Poisson brackets. Thus it is necessary to work with Dirac brackets, interpreting $\pi_i\approx 0$ as a pair of second class constraints \cite{GJPP}. The Poisson algebra among this pair is,
 \begin{equation}
    C_{ij}=\{\pi_i, \pi_j\}= -eB\epsilon_{ij}
    \label{matrix}
 \end{equation}
 Now the Dirac brackets (denoted by a star) among the coordinates is obtainable using the definition,
 \begin{equation}
     \{x_i, x_j\}^*=\{x_i, x_j\} - \{x_i, \pi_k\}C^{kl}\{\pi_l, x_j\}
     \label{db}
 \end{equation}
 where $C^{kl}= (eB)^{-1}\epsilon^{kl}$ is the inverse of (\ref{matrix}). The result is,
 \begin{equation}
     \{x_i, x_j\}^*=\frac{1}{eB} \epsilon_{ij}
     \label{landaualgebra1}
 \end{equation}
 which reproduces (\ref{landaualgebra}). Thus, given a hamiltonian like (\ref{lanham}) it is possible to compute the relevant Dirac brackets by this approach, even if the lagrangian is not known.

 It is now feasible to make contact with the volume preserving diffeomorphisms satisfying the nonlinear closure (\ref{nonlinear}) and (\ref{nonlinearparameter}) discussed in the previous section. The algebra of the parameters (\ref{nonlinearparameter}) is now lifted to a commutator,
 \begin{equation}
   [\eta(x), \rho(x)]= il^2 \epsilon_{ij}\partial_i\eta\partial_j\rho
   \label{commutator}
 \end{equation}
 which may be expressed in terms of the algebra of the coordinates as,
 \begin{equation}
     [\eta(x), \rho(x)]=\partial_i\eta\partial_j\rho[x_i, x_j]
     \label{palgebra}
 \end{equation}
Comparing (\ref{commutator}) and (\ref{palgebra}) yieds,
 \begin{equation}
     [x_i, x_j]= il^2 \epsilon_{ij}
     \label{palgebra1}
 \end{equation}
 Identifying,
 \begin{equation}
     l^2=\frac{1}{eB}
     \label{identification}
 \end{equation}
 the parametric algebra associated with the volume preserving diffeomorphisms becomes identical with the algebra (\ref{landaualgebra}){\footnote{The classical bracket is lifted to a commutator by multiplying it with $i$.}} of the lowest Landau level problem.
 
 \subsection{Physics of lowest Landau level problem and algebra of charges}
 
 The nontrivial charge algebra (\ref{chargealgebra}) is a characteristic of noncommuting coordinates. If the coordinates were commuting, the charge algebra would be trivial, i.e. vanishing. Indeed such a noncommutative  algebra has appeared naturally  in the context of noncommutative fluid dynamics and  magnetohydrodynamics.  Moreover, since fluid dynamics can be interpreted as an example of a volume preserving diffeomorphism invariant theory, it is possible to understand the relation (\ref{chargealgebra}) from that point of view.
 
 In the hamiltonian formulation of Eulerian fluids, the particle coordinate is denoted by $\bar X_i(t)$ where $i$ labels the particle. Then the charge density is given by,
 \begin{equation}
     \rho(t, \bar x)=\Sigma_{i=1}^{i=N}\,\delta
    ( \bar X_i(t)-\bar x)
     \label{fluid}
 \end{equation}
 where, for simplicity, the mass parameter has been set to unity and $N$ is the number of particles. The discrete particle labels may be replaced by continuous spatial arguments (omitting time),
 \begin{equation}
     \rho (\bar x) =\int_{volume} \delta(\bar X(x)-\bar x)
     \label{cont}
 \end{equation}
 A volume integral of the density $\rho$ yields the total mass which has been normalised to unity. If the coordinates commute the charge algebra vanishes. However, if we take the algebra among the coordinates that is relevant for the  lowest Landau level problem (\ref{landaualgebra1}), so that,
 \begin{equation}
     [X_i(\bar x), X_j(\bar y)]=\frac{i}{eB} \epsilon_{ij}\delta (\bar x- \bar y)
 \end{equation}
 which is the field theoretic analogue of (\ref{landaualgebra1}), 
 lifted to a commutator, we obtain \cite{GJPP},
 
 \begin{equation}
     [\rho(\bar x), \,\rho(\bar y)]= \frac{i}{eB}\epsilon^{kl}\partial_k\rho(\bar x)\partial_l\delta(\bar x- \bar y)
     \label{newchargealgebra}
 \end{equation}
 which reproduces (\ref{chargealgebra}) after the identification (\ref{identification}) is used.
 
 It is also possible to construct noncommutative magnetohydrodynamics such that the cherished charge algebra  (\ref{chargealgebra}) or (\ref{newchargealgebra}) is obtained. This  has a close parallel with the physics of the lowest Landau level problem including the corresponding Dirac analysis.
 
 The equations governing the motion of a charged fluid with density $\rho$ and mass parameter $m$ (introduced for dimensional purpose) moving on a plane with velocity $\bar v$, subjected to a constant external magnetic field $B$ perpendicular to the plane, are given by the continuity equation,
 \begin{equation}
     \dot \rho + \partial_i(\rho v_i) = 0
     \label{continuity}
 \end{equation}
 and the Euler equation,
 \begin{equation}
     m \dot v_i + mv_j\partial_j v_i =e \, \epsilon_{ij} v_j B +F_i
     \label{euler}
 \end{equation}
 where extra forces $F_i$ are defined from a potential \cite{GJPP},
 \begin{equation}
     F_i(\bar x) = -\partial_i \frac{\delta}{\delta \rho (\bar x)} \int d^2 x \, V
     \label{potential}
     \end{equation}
 
 The continuity and Euler equations (\ref{continuity}), (\ref{euler}) are obtained by taking the Poisson brackets of $\rho$ and $v_i$ with the hamiltonian,
 \begin{equation}
     H= \int d^2x\, \Big(\rho \frac{\pi_i^2}{2m} + V\Big) \,\,; \,\,\, (\pi_i = m v_i)
     \label{hamiltonianmhd}
 \end{equation}
 provided the brackets among the basic variables are taken as,
 \begin{eqnarray}
  \{\rho(\bar x), \rho(\bar y)\} &=& 0 \cr
  \{\pi_i(\bar x), \rho(\bar y)\} &=& \partial_i\delta(\bar x-\bar y)\cr  
  \{\pi_i(\bar x), \pi_j(\bar y)\} &=& -\epsilon_{ij} \frac{1}{\rho} \Big(m \omega(\bar x) + eB\Big) \delta(\bar x-\bar y)
  \label{fundamental}
 \end{eqnarray}
 where,
 \begin{equation}
     \epsilon_{ij}\omega = \partial_i v_j - \partial_j v_i
     \label{vorticity}
 \end{equation}
 is the vorticity of the fluid.
 
 For a strong magnetic field the mass parameter goes to zero as may be seen from (\ref{euler}). In that case, for a meaningful hamiltonian (\ref{hamiltonianmhd}) to exist, the momenta $\pi_i$ should vanish. A more clear cut justification for this was given in the basic quantum mechanical Landau problem.{\footnote{See the discussion below (\ref{discussion}).}}
 Such a lagrangian is nonexistant here but the hamiltonian has a similar structure. Putting $\pi_i=0$ directly in the above algebra leads to inconsistencies. Hence recourse is taken to the Dirac analysis of constraints. The constraint $\pi_i\approx 0$ is implemented weakly. In fact it forms a pair of second class constraints. These may be strongly imposed by calculating the relevant Dirac brackets. The $\rho-\rho$ Dirac bracket, elevated to a commutator, is precisely (\ref{newchargealgebra}) \cite{GJPP}. Using this algebra the appropriate equations of motion are reproduced by taking the hamiltonian as,
 \begin{equation}
     H=\int d^2 x\, V
 \end{equation}
 obtained by putting $\pi_i^2=0$ in (\ref{hamiltonianmhd}),{\footnote{For an alternative viewpoint regarding the formulation of noncommutative magnetohydrodynamics and, especially, the derivation of the charge algebra (\ref{newchargealgebra}), see \cite{RB}. It is based on 't Hooft's \cite{G} notion of information loss and dissipation in quantum mechanics.}}
 
 \section{Conclusions}
 We have given a hamiltonian analysis of higher rank symmetric gauge theories, focusing on aspects that were either partially or, not highlighted. Instead of introducing constraints and transformation laws by hand, we proceed from a higher derivative lagrangian and generate these by adopting Dirac's algorithm of constrained systems. In this way we do not miss any constraints, either first or second class. Neither is there any lack of uniqueness or consistency. 
 
 Of particular interest is the $(2+1)$ dimensional traceless scalar charge theory which was treated here in a different way by first imposing the traceless condition $A_{ii}=0$ in the lagrangian by means of a multiplier and then doing the canonical analysis.  This is important since the canonical analysis and imposition of the traceless constraint are noncommutative, as explained in details below (47).  The theory led to first and second class constraints, both of which have distinct roles. Solving the traceless constraint explicitly by expressing $A_{ij}$  in terms of another (traceful) field $(h_{ij})$, it was found that the new theory was equivalent to linearised gravity with volume preserving diffeomorphisms. Explicit forms for the action, constraints and the transformations rules were found in the theory describing linearised gravity. A direct connection of this symmetry with that in the lowest Landau level problem was shown. The modified charge algebra was identical to that found in noncommutative fluid dynamics or in magnetohydrodynamics in presence of a strong magnetic field.
 
 
 The systematic analysis of constraints done here may be extended in other directions. One possibility is the inclusion of higher derivative Chern-Simons terms and study their effects. In standard gauge theories their inclusion has led to many interesting results in odd dimensions and one might be hopeful here also. Especially significant would be to investigate duality and dual transformations where Chern-Simons terms have a useful role. The existence of more than one Chern-Simons type term  makes matters more interesting. The surprising connections of these higher derivative theories with the physics of volume preserving diffeomorphisms encompassing the lowest Landau level problem, noncommutative fluid dynamics and magnetohydrodynamics, to name a few, show the potential importance of higher rank symmetric tensor theories in other areas besides the usual ones like defects in solids, supersolids, spin liquids, superfluid vortices etc. where they have already proved  their utility.
 
 \bigskip
 
 \noindent {\bf {Acknowledgements}}
 
 This work was supported by  a (DAE) Raja Ramanna Fellowship. I also thank the Director, S.N. Bose Centre, for providing  necessary facilities.

 \end{document}